\documentclass[12pt]{article}
\usepackage{a4,amsmath,amsxtra,amssymb}

\begin{document}

\baselineskip=1.5\baselineskip

\centerline{\bf{\Large A HYPOTHESIS ON PRODUCTION OF TACHYONS}}

\vskip36pt

\centerline{\bf J. K.
Kowalczy\'nski\footnote[1]{E-mail address: jkowal@ifpan.edu.pl}}

\vskip12pt

\centerline{\it Institute of Physics, Polish Academy of Sciences,}

\centerline{\it Al. Lotnik\'ow 32/46, 02--668 Warsaw, Poland}

\vskip84pt

\noindent
An exact solution of the Einstein--Maxwell equations
yields a general relativistic picture of the
tachyonic phenomenon, suggesting a hypothesis on the tachyon
creation. The
hypothesis says that the tachyon is produced when a neutral and
very heavy (over 75~GeV/$c^{2}$) subatomic particle is placed in
electric and magnetic fields that are perpendicular, very strong
(over $ 6.9 \times 10^{17}$~esu/cm$^{2}$ or oersted),
and the squared ratio of their strength lies in the interval~(1,5].
Such conditions can occur when nonpositive subatomic particles of
high energy strike atomic nuclei other than the proton.
The kinematical relations for the produced tachyon are given.
Previous searches
for tachyons in air showers and some possible causes of their
negative results are discussed. Experiments with the use of the
strongest colliders and improvements in the air shower experiments
are suggested. An unfortunate terminology is
also discussed.

\vskip12pt


\newpage

\centerline{{{\bf\large 1 Introduction}}}

\vskip12pt

\noindent
The long-lasting discussion on the tachyonic causal paradoxes
has yielded a large number of self-contradictory publications,
which has caused a cautious attitude of many physicists towards the
tachyon. The problem of these paradoxes has lucidly been reviewed
by Girard and Marchildon [1] (though in fact I disagree with some
of their conclusions), and the essence of construction of
the known paradoxes has thoroughly been analyzed in Ref.~[2]. A
large part of the most representative literature of the subject is
cited in Refs. [1,2] (see also the end of Footnote~14). It has been
concluded that the problem of whether the paradoxes may be
eliminated within the standard theory of relativity remains still
open (see, however, the end of the paragraph next but one), and
that there exist such consistent extensions of this theory in which
the known paradoxes are eliminated. The latter conclusion means
that there is no contradiction between relativity and the tachyon's
existence, though today we do not yet know whether the tachyon
exists in nature.

The discussion on tachyons has been conducted mainly at the
special relativity level with its standard poor pictures of the
tachyonic phenomenon. In these pictures the tachyon does not
generate any field. In general relativity the situation is
different, since there we know some exact solutions of the
Einstein and
Einstein--Maxwell equations that describe spacetimes generated by
the tachyonic sources. These spacetimes, filled with gravitational
and electromagnetic fields, are bounded by tachyon shock waves which
are singular in terms of these solutions. Creation of the tachyon
shock wave occurs also in a quantum description of the tachyon's
motion~[3].\footnote[2]{In Ref.~[3]
there is a misprint. Namely, Eq.~(22)
should read $F = a_{-1}\int Md\zeta$ (notation after Ref.~[3]).}
It is interesting that this
description includes certain tachyonic four-momentum relations that
agree with the general relativistic pictures of the tachyonic
phenomenon but do not agree with the special relativistic ones. In
sum, our present-day knowledge of the tachyon strongly suggests
that special relativity is too confined to describe tachyons (in
classical terms), and that at least general relativity is
necessary.

In fact, one of the exact tachyonic solutions seems to be of
special importance for the problem of tachyons and for our
hypothesis. This solution is presented in Section~2.
It differs from the rest of the known tachyonic solutions in two
properties: first, it has neither a bradyonic nor a luxonic
counterpart, i.e. it is a {\it specifically tachyonic\/} solution;
and second, it has no independent term which would include a
masslike quantity.\footnote[3]{We do
not know what the counterparts of the bradyonic mass and/or charge
mean in the tachyonic formulae (an example is given in Section~6),
since we do not have any operational definitions of such quantities.
I have
therefore proposed to use the terms ``masslike quantity" [3,4] and
``chargelike quantity" [4] for these counterparts. (The terms
``pseudo-mass(-charge)'' or ``quasi-mass(-charge)'' are shorter
but semantically inferior.)
In the tachyonic
literature it is stated, from time to time, that the subluminal
electric (magnetic) charge becomes, or behaves like, a magnetic
(electric) charge when it becomes superluminal. So far, however,
there is no operational model for this statement.} The second
property is
important for our hypothesis and is discussed at the beginning of
Section~3. If we assume the picture of the tachyonic phenomenon
resulting from
this solution, i.e. a picture obtained {\it within\/} standard
relativity, then the construction of the known paradoxes
becomes questionable~[4].

Various experimental searches for ionizing tachyons have been
described in a number of papers. A large majority of them is
cited in Refs.
[5--10]. The experiments were of low and high energy type. Failure
of the low energy experiments is explicable by our hypothesis, as
will be seen in Sections 4 and 5. In the high energy experiments
air showers were exploited; and many of the experiments
have reported detection
of tachyon candidates but as statistically insignificant data. A
single possibly positive result [11] has also been rejected~[5].
This situation has presumably disheartened most experimenters (the
last relevant record in the Review of Particle Properties [9] is
dated 1982~[8]), though some efforts were still made~[10].
According to our hypothesis, however, air shower (and accelerator)
experiments may
be successful and they are discussed in Section~5.
Though the tachyons considered in this paper are
ionizing objects,
experiments yielding tachyonic~(?) neutrinos are briefly commented
in Section~6, where also an unfortunate terminology is
criticized.

\vskip16pt

\centerline{{{\bf\large 2 The solution}}}

\vglue12pt

\noindent
The basis of our hypothesis is an exact solution of the
current-free Einstein--Maxwell equations
$$
 G_{\mu\nu} = 2c^{-4}\kappa\left(F_{\rho\mu}F_{\nu}\vphantom{F}^{\rho}
 + \textstyle{1\over4}g_{\mu\nu}F_{\rho\tau}F^{\rho\tau}\right),
$$
$$
 F_{\left[\mu\nu,\rho\right]} = 0,\qquad F^{\mu\nu}
 \vphantom{F}_{;\nu} = 0,
$$
where $ G_{\mu\nu}$, $ F_{\mu\nu}$, and $ g_{\mu\nu}$ are the
Einstein, electromagnetic field, and metric tensors, respectively,
$ c$ is the speed of light in vacuum, and
$ \kappa $ is the Newtonian gravitational constant. The solution in
question is as follows:
\begin{align}\label{7} %label (1)-(7)
  & ds^{2} = ds_{0}^{2} + ac^{-4}{\kappa}p^{-1}\left(2\theta +
  {\textstyle{1\over2}}\ln
  |q| - p^{-1}q\right)dq^{2}\/, \\
  & ds_{0}^{2} := p^{2}\left(d\theta^{2} +
  \text{e}^{-2\theta}d\phi^{2}
  \right) + 2\,dp\,dq + dq^{2}\/, \\
  & aq \geq 0\/, \\
  & F_{\phi\theta} = -\chi \text{e}^{-\theta},\qquad F_{\phi q} =
  {\textstyle{1\over2}}\chi
  q^{-1}\text{e}^{-\theta},\qquad F_{\theta q} =
  -{\textstyle{1\over2}}\varepsilon
  q^{-1},\nonumber \\
  & F_{pq} = -\varepsilon p^{-2},\qquad F_{\phi p} = F_{\theta p}
  = 0\/, \\
  & \chi^{2} + \varepsilon^{2} = aq\/, \\
  & F_{\mu\nu}F^{\mu\nu} = 2p^{-4}\left(\chi^{2} -
  \varepsilon^{2}\right)\/, \\
  & F_{\mu\nu}\widetilde{F}^{\mu\nu} = -4p^{-4}\chi\varepsilon\/,
\end{align}
where $ \phi $ and $ \theta $ are dimensionless coordinates, $ p $
and $ q $ are coordinates having the length dimension, $ a $ is an
arbitrary constant having the energy dimension, and
$ \widetilde{F}^{\mu\nu} $ is the dual of $ F_{\mu\nu}$.
All these quantities are real.

The form $ ds_{0}^{2} $ is the flat part of form~(1). Inequality
(3) is a condition of solvability of the Einstein--Maxwell equations
in the case under consideration. The metric form (1)--(3) describes
more than one spacetime. Each of the spacetimes has boundaries $
S_{p} $ and $ S_{q}$, where $ S_{p} $ is determined by relations $
p = 0 $ and $ aq \geq $ 0, and $ S_{q} $ by $ q = 0 $ with a limit
$ p = 0 \,\cap\, q = $ 0. These spacetimes can be extended neither
through $ S_{p} $ nor $ S_{q}$, since each of the conditions $ p =
0 $ and $ q = 0 $ determines the strongest curvature singularity of
our solution, namely a singularity (infinite value) of $
R_{\mu\nu\sigma\tau}R^{\mu\nu\sigma\tau} $ and of $
R_{\mu\nu\sigma\tau}R^{\sigma\tau\omega\kappa}R_{\omega
\kappa}\vrule width0pt^{\mu\nu}
$. Every two-dimensional surface determined by conditions~(1), (2),
$ p = {\rm constant} \neq 0$, and $ q = {\rm constant}$ has the
negative Gaussian curvature. This and the fact that our solution
belongs to the Robinson--Trautman class [12] mean that the metric
form (1)--(3) describes spacetimes generated by tachyons
[13--15].\footnote[4]{Solutions describing
gravitational waves also belong to the Robinson--Trautman class
[12,13]. It is therefore interesting from the psychological point
of view that the problem of gravitational waves is considered as
very important whereas some physicists consider that the problem of
tachyons cannot be treated seriously,
though both phenomena have the same empiric
status: they are not yet confirmed. Massive experiments to search
for gravitational waves have been performed and very expensive ones
are planned, while the experimenters searching for tachyons have
been very modestly equipped.} The geometric standards
of recognition of the solution under
consideration are given in Ref.~[15]. In Ref.~[4] our solution is
referred to as $ \Omega_{1}$.

Formulae (1)--(7) are simple but they do not depict the physical
situation. After making the coordinate transformation
\begin{align}\label{} %label (8)
  & \phi = y\left(T - x\right)^{-1},\quad \theta =
  {\textstyle{1\over2}}\ln
  \left(T^{2} - x^{2} - y^{2}\right) - \ln \left(T -
  x\right),\nonumber \\
  & p = j\left(T^{2} - x^{2} - y^{2}\right)^{1/2},\qquad q = Z -
  p,\nonumber \\
  & T \geq \left(x^{2} + y^{2}\right)^{1/2} \geq 0,\qquad j = \pm1,
  \quad ja < 0, \quad jp \geq 0,\nonumber \\
  & Z := \gamma\left(z - vt\right), \qquad T := \gamma\left(ct -
  c^{-1}vz\right),\nonumber \\
  & \gamma := \left(1 - c^{-2}v^{2}\right)^{-1/2} \geq 1, \qquad |v|
  < c,
\end{align}
where $ v $ is a transformation parameter having the speed
dimension, Eqs. (1) and (4) explode, but from Eq.~(2) we get a
familiar form
\begin{equation}\label{9}
  ds_{0}^{2} = dx^{2} + dy^{2} + dz^{2} - c^{2}dt^{2}. %(9)
\end{equation}
In terms of the obtained coordinate system $ x$, $ y$, $ z$, $ t
$ we can explicitly describe the situation both in spacetime and in
space, and we can reveal a property of our electromagnetic field $
F_{\mu\nu} $ important in the contex of our hypothesis; and
this is done in brief just below.

In spacetime the boundary $ S_{p} $ is a semi-infinite light wedge.
Its edge is a semi-infinite spacelike line $ x = y = T = 0 $ which
is the world line of the tachyon generating each one of the
spacetimes (1)--(3). The boundary $ S_{q} $ is a fragment of
the light
cone. In the case under consideration these two boundaries are
smoothly\footnote[5]{We
take here into account the expanding $\left(T \geq 0\right)$
and convex $\left(ja < 0\right)$ spacetimes since only such a type
of spacetimes (1)--(3) can be real and autonomous [4,16]. References
[2,16] are commented in Appendix A in Ref.~[4].} tangent and form a
null hypersurface $ S = S_{p} \cup S_{q}$
enveloping the generated spacetime. The beginning of
the edge and the vertex
of the light cone coincide at a spacetime point (event) which can
therefore be interpreted as a {\it creation point\/} of the tachyon
and,
consequently, of the whole tachyonic phenomenon considered here.
The existence of this geometrically distinguished event is an
invariant property of our solution and makes a reasonable physical
interpretation possible. Transformation (8) was chosen so as
to have $ x = y = z = t = 0$ at this event.

In space we have a surface consisting of two parts, conical
with axis $ z $ ($S_{p} $ in space) and spherical
with centre $ x = y = z = 0$ ($S_{q} $ in space), which
are smoothly tangent.
This surface expands along its normals with the speed of light. In
consequence, the vertex of the cone moves along a semi-axis $ z $
with a constant velocity $ w $ such that
\begin{equation}\label{10}
  vw = c^{2}.
\end{equation}
Thus $ |w| > c$, i.e. we have a {\it pointlike tachyon}. The
spherical part can be interpreted as a shock signal of a birth at
the point $ x = y = z = 0 $ and instant $ t = 0$, and the conical
part as a shock wave of the born tachyon. Since these two parts are
smoothly tangent, the picture of the whole
phenomenon is quite realistic. This picture is the most realistic
one among the
general relativistic pictures of the tachyonic phenomenon known
today, and it is probably the simplest realistic picture obtainable
within general relativity.

The infinite curvature and electromagnetic field on the null
hypersurface $ S $ (by relations (1)--(7) and the condition $ p =
0$ or $ q = 0$), and thus on the shock surface in space, are of
course mathematical exaggerations frequently occurring in
theoretical descriptions of nature. In reality there is a thin
``skin'' enveloping the spacetime (space) generated by the tachyon.
This ``skin'' is made of finite but relatively strong fields --
gravitational and electromagnetic. The presence of the
electromagnetic field means that our tachyon is an {\it ionizing\/}
object.

The subject-matter of the three preceding paragraphs is
discussed wider
in Ref.~[16] and much wider in Ref.~[4]. The tachyonic phenomenon
under consideration is depicted in various reference frames
by figures in Refs. [4,16].

When the electromagnetic field (4) and (5) is investigated in terms
of the coordinate system $ x$, $ y$, $ z$, $ t$, it appears that
there exists a part independent of $ x$, $ y$, and $ z$. In the
quasiflat case ($ ds^{2} \cong ds_{0}^{2} $),
considered in the further text, $ x $, $ y $, and $
z $ are spacelike coordinates (see Eq.~(9)), i.e. we have then a
background part of our electromagnetic field. The existence of this
part is one of the guides to our hypothesis. Details are given
in Ref.~[4].

\vskip16pt

\centerline{{{\bf\large 3 Premises of the hypothesis}}}

\vglue12pt

\noindent
The creation point of the tachyon is singular in terms of our
solution (see Section~2), and therefore the conditions of
production of the tachyon cannot be calculated within the exact
theory based on this solution. The calculation of these conditions
needs some additional assumptions, e.g. that regarding
the finite strength of the fields present
on $ S $ (see the last but two paragraph in
Section~2). Though these assumptions are not contradictory to our
solution, we speak here of a hypothesis only and not of a
theory.

The known tachyonic solutions of the Einstein--Maxwell equations
different from our solution, as well as their luxonic and bradyonic
counterparts, include terms containing a masslike quantity
(mass in the bradyonic solutions; see Footnote~3). These terms are
independent of the electromagnetic ones and therefore each of them
can be removed only by virtue of our arbitrary assumption. From
relations (1)--(5) we see that our solution does not include such a
term. This is an essential property of the metric form (1) and~(2).
In fact, for this form such a term is additive and reads
$ 2m_{0}c^{-2}{\kappa}p^{-1}dq^{2} $ [4,14,15], where $ m_{0} $ is a
constant masslike quantity, but for $ a \neq 0 $ the coordinate
transformation $ \theta \to \theta -a^{-1}m_{0}c^{2} $ and $ \phi
\to \phi\exp \left(-a^{-1}m_{0}c^{2}\right) $ annihilates this term
and restores the form (1) and~(2). In our case therefore the
gravitational field, i.e. the direct cause of spacetime curvature,
does not exist autonomously but is generated by the electromagnetic
field (4) and~(5). The factor $ c^{-4}{\kappa} \approx 10^{-49} $
g$^{-1}$cm$^{-1}$s$^{2} $ (see Eq.~(1)) is, however, so small that
even if the field (4) and (5) were by many orders of magnitude
stronger than the strongest electromagnetic fields observed so far,
the spacetime curvature would be completely negligible.
Thus, even for a very strong field
(4) and (5), our spacetime is practically flat everywhere,
$ ds^{2} \cong ds_{0}^{2} $,
including the ``skin'' (see the last but two
paragraph in Section~2). This means that our solution is proper to
describe an ionizing tachyon {\it belonging to the microworld}.
(From time to time general relativity directly enters the
microworld; see, e.g., Section~7 in Ref.~[17].) When passing to the
flat spacetime and microworld, our picture of the tachyonic
phenomenon is preserved, since in virtue of relations (3)--(7) our
electromagnetic field is (formally) infinite everywhere on the
boundary $ S $ (as $ p = 0$ or $ q = 0$ on~$ S $).

Equation (5) is analogous to $ \chi_{0}^{2} +
\varepsilon_{0}^{2} = b^{2} $, where $ b $ is an electromagnetic
constant occurring in the well-known Reissner--Nordstr\"om (R--N)
solutions of the Einstein--Maxwell equations. In the bradyonic R--N
solution constants $ \chi_{0} $ and $ \varepsilon_{0} $ are
charges of magnetic and electric monopoles, respectively, and in
the tachyonic R--N solution they are chargelike quantities of
monopoles of indefinite meanings (see Footnote~3). Thus the case $
\chi_{0} = 0 $ and $ \varepsilon_{0} \neq 0 $ and the case $ \chi_{0}
\neq 0$ and $ \varepsilon_{0} = 0 $ are pure cases in which only
one type of charges or chargelike quantities occurs.
Considering the analogy
just mentioned, we have a pure case when $ \chi = 0 $ and $
\varepsilon \neq 0 $ or when $ \chi \neq 0 $ and $ \varepsilon =
0$. By Eq.~(7) in each of these two cases the electric and
magnetic fields are perpendicular everywhere. (This takes also
place in the pure cases of the R--N solutions.) The tachyon
generating the field (4) and (5) with $ \chi = 0 $ and $
\varepsilon \neq 0 $ will be called the {\it e-tachyon\/}
(electric type tachyon; predominance of
the electric field since $ F_{\mu\nu}F^{\mu\nu} < 0$, see Eq.~(6)),
and that with $ \chi \neq 0 $ and $ \varepsilon = 0$ will be called
the {\it m-tachyon\/} (magnetic type tachyon; predominance
of the magnetic field since $
F_{\mu\nu}F^{\mu\nu} > 0$, see Eq.~(6)). Note that nothing
is said about the chargelike quantities of these tachyons.

On the analogy of the subluminal microworld, in
which only one type of charges (electric) is known, we may suspect
that only one type of our tachyons exists in
nature (i.e. either the e-tachyons or the m-tachyons),
but today we do not yet know which one. Thus, for safety, both
types should be considered. Note that the existence
of mixed cases (our $ \chi\varepsilon \neq 0$, $
\chi_{0}\varepsilon_{0} \neq 0 $ of~R--N called dyon in the bradyonic
case) seems unnatural when no pure case exists autonomously.

It is known that, in terms of relativity, no tachyon can be at
rest\footnote[6]{It has been shown
in terms of the invariant properties of the light cone~[2] (and
less precisely but in a simpler and shorter way in Ref.~[18]) and
in terms of the group theory [19] that the concept of superluminal
reference frame (i.e. the frame in which a tachyon may be at rest)
does not exist in relativity, and that every consistent extension
of relativity by adding this concept yields a notional system
unacceptable from the physical point of view. Unfortunately, an
extensive literature exists in which superluminal frames and
transformations are seriously treated in the context of relativity
(cf. Footnote~15).} (i.e. every tachyon is always in motion and
therefore it determines a direction in space), and that there is no
invariant (with respect to all the time-irreversible Lorentz
transformations) past-future orientation along the tachyon's world
line. Besides, in our case the event of tachyon's birth and the
spacelike orientation along the tachyon's world line are
determined, owing to the existence of the creation point in our
solution (see Section~2). In contrast, the flat spacetime (being
now the arena of our considerations, see the second paragraph of
this section) includes the past-future orientation and its space is
believed to be homogeneous and isotropic. Thus
the tachyon should be ``informed'' already {\it in statu nascendi\/}
of its properties
just mentioned, to ``let him know'' how to come into being in our
space of undistinguishable points and directions. Such
``information'' can, however, be introduced into this space only by
creating proper physical conditions. In our case it is most
natural to have an electromagnetic field which will coincide
with the background\footnote[7]{We
have here an analogy to the wave-particle duality of the subluminal
microworld. Namely, nonlinear electrodynamics describes
faster-than-light electromagnetic signals which, however, must have
a background electromagnetic field to propagate [20,21].} part
of the field generated by the tachyon (see the end of
Section~2), and a material micro-object immersed in this field.
Such a micro-object determines the place of the tachyon's birth
(creation point demanded by our solution), and the electromagnetic
field indicates the direction and sense of the tachyon's motion.
Further these micro-object and electromagnetic field are called the
{\it generative particle\/} and the {\it initiating field}.

The production conditions mentioned just above are kinematical and
should be supplemented with the strength of the initiating field and
with the
information about the generative particle. We can do this by using
the Heisenberg time-energy uncertainty relation. The combining of
this relation, fundamental in quantum physics, with our classical
description of the tachyonic phenomenon seems to be a proper
move since we deal here with a tachyon belonging to the
microworld. This combination and the following
procedures, simple or involving laborious calculations, are
presented in detail in Ref.~[4]. Here we present only their
results. It appears that the initiating field must be very
strong,\footnote[8]{We have here an
analogy to the spontaneous creation of bradyonic particles in very
strong electromagnetic fields (for review see, e.g., Ref.~[22]).
The minimal strength of these fields is by only one order of
magnitude smaller than that of our initiating field (given by
relations (11)--(13) and~(19)). The essential difference consists in
that those bradyons are created in vacuum whereas our
tachyon in the generative particle.} and that the generative
particle must be a neutral subatomic particle of very large rest
mass (inequality~(20)). This mass is an additional fuel required by
the energy conservation law for producing the tachyon. Our
hypothesis says nothing about other properties of the generative
particle, e.g. quantum numbers. We may assume that depending
on the situation some additional entities may be produced, e.g. if
the proper conservation laws
hold.\footnote[9]{This problem is
discussed in Footnote 26 in Ref.~[4]. Let us supplement that
footnote by noting that the simultaneous production of tachyonic
neutrinos (if they exist, see Section~6) would be an interesting
possibility.}

\vskip16pt

\centerline{{{\bf\large 4 The hypothesis}}}

\vglue12pt

\noindent
The hypothesis says that the tachyon is produced when a neutral
subatomic particle of sufficiently large rest mass (the generative
particle) is placed in the strong electromagnetic
field (the initiating field) described just below. The generative
particle is then annihilated giving birth to the tachyon.

In this section we use the Lorentzian coordinate system introduced
in Section~2 (see Eq.~(9)). According to Sections 2 and 3 the {\it
proper\/} reference frame of the generative particle can be endowed
with this coordinate system in such a way that the generative
particle is at the origin $ x = y = z = 0 $ of the spacelike
coordinates. In this section all quantities, relations, and
situations are presented in terms of this reference frame.

Let $ {\bf E} $ and $ {\bf H} $ be accordingly the electric and
magnetic three-vectors of the initiating field, and let their
components be denoted by $ E_{i} $ and $ H_{i} $. In order to
produce the tachyons under consideration we should have the
following two types of the initiating field:
\begin{align}\label{11} %label 11
  & E_{x} = \mp\gamma\lambda\Xi,\qquad E_{z} = \pm2j\lambda\Xi,
  \qquad H_{y} = \mp j\gamma\lambda\Xi,\nonumber \\
  & E_{y} = H_{x} = H_{z} = 0,
\end{align}
in which the e-tachyon is produced, and
\begin{align}\label{12} %label 12
  & E_{y} = \pm\gamma\lambda\Xi,
  \qquad H_{x} = \mp j\gamma\lambda\Xi,
  \qquad H_{z} = \pm2\lambda\Xi,\nonumber \\
  & E_{x} = E_{z} = H_{y} = 0,
\end{align}
in which the m-tachyon is produced, and where
$$
\lambda := \left(\gamma^{2} + 4\right)^{-1/2} > 0,\qquad \Xi > 0,
\eqno (13)$$ and $ j $ is determined by relations~(8). The tachyon
produced in the generative particle and fields (11)--(13) will be
moving along a semi-axis $ z $ with a velocity $ w $ such that
$$
jw < 0,\eqno (14)
$$
where $ w $ is related to $ \gamma $ by relations (8) and~(10).

From relations (11)--(13) we see that
$$
{\bf E} \perp {\bf H},\qquad |{\bf E}| \neq |{\bf H}|,\qquad |{\bf
E}||{\bf H}| \neq 0,\eqno (15)
$$
and that $ \Xi = |{\bf E}| > |{\bf H}| $ in the case (11) and $ \Xi
= |{\bf H}| > |{\bf E}| $ in the case~(12).

Let $ U $ be defined as follows: $ U = |{\bf H}|^{-1}|{\bf E}| $ in
the case (11) and $ U = |{\bf E}|^{-1}|{\bf H}| $ in the case~(12).
Thus, by relations (11)--(13), we have $ U > 1 $ and
$$
U^{2} = 1 + 4\gamma^{-2} = 5 - 4c^{2}w^{-2},\eqno (16)
$$
i.e.
$$
1 < U^{2} \leq 5.\eqno (17)
$$
Note that in accordance with the known properties of the spacelike
world lines we may have $ |w| = \infty $.
If the angle between the tachyon path (a
semi-axis $ z $) and the longer three-vector of the initiating
field is denoted by $ \alpha $, then
$$
\sin\!\alpha = U^{-1}.\eqno (18)
$$

By generating perpendicular electric and magnetic fields we determine
empirically the directions in space. If these fields satisfy the
condition~(17), then, according to the hypothesis, for each type of
tachyons under consideration Eqs. (16) and (18) determine four
variants of the complete kinematical conditions for the produced
tachyon. The existence of four variants results from relations
(11)--(14) and~(18). Namely, there are double signs of the nonzero
components $ E_{i} $ and $ H_{i} $, a double sign of $ j $ (i.e. a
double sign of $ w $ since $ jw < 0$), and $ \sin\!\alpha =
\sin\!\left(\pi - \alpha\right) $, i.e. we apparently have eight
variants, but each one of these three items depends on two
others.

In order to determine the principal empiric conditions for the
production, we should also know the quantity $ \Xi $ and the rest
mass $ M $ of the generative particle. By using the Heisenberg
time-energy uncertainty relation (cf. the end of Section~3) we can
estimate the lower limits of $ \Xi $ and $ M $.

In the case of $ \Xi $, we fairly easily [4] get
$$
\Xi \gtrsim 6.9 \times 10^{17} {\rm\ esu/cm}^{2}
{\rm\ or\ oersted.}\eqno (19)
$$

In the case of $M$, I am able to estimate its lower limit only when
$|w| \cong c$ (thus for $U \cong 1$; note that $|w| > c$ and $U >
1$), i.e. when the produced tachyon is very ``slow'' in the proper
reference frame of the generative particle.\footnote[10]{Such a
tachyon can, however,
be observed as considerably faster than light if the sense of its
velocity is opposite in the laboratory reference frame to the sense
of the generative particle velocity (sufficiently high but subluminal
of course); cf. remarks on the backward tachyons in Section~5.}
Laborious calculation [4] gives
$$
M \gtrsim 75\ {\rm GeV/}c^{2}.\eqno (20)
$$

Our hypothesis concerns the production of the tachyons for which
the hypersurfaces~$ S $ (see Section~2)
are convex; and such tachyons can
exist autonomously. Let us call them {\it principal tachyons}. Each
principal tachyon may be accompanied with an arbitrary (formally)
number of tachyons for which the hypersurfaces $ S $
are concave. The latter
tachyons cannot exist autonomously but they can exist if they form
a ``star of tachyons'' together with a principal
tachyon. Let us call them {\it accompanying tachyons}.
All the tachyons
forming their ``star'' are born at one event (common creation point,
for details see Refs. [4,16]).

\vskip16pt

\centerline{{{\bf\large 5 Comments on the empiric possibilities}}}

\vglue12pt

\noindent
The production conditions determined by our hypothesis can occur in
high energy collisions with atomic nuclei other than the
proton. In such collisions we can locally obtain the conditions~(15)
(for details see Ref.~[4]) and the relativistic
intensification of the electromagnetic fields of nuclei necessary
to satisfy the condition~(19). It is easy to calculate that this
intensification gives $ U \cong 1$, i.e. the condition (20) holds.
Thus the gauge boson Z$^{0} $ is the lightest known candidate for
the generative particle. Though the mean life of this boson is very
short, the production conditions can be satisfied. In fact, if a
subatomic particle of sufficiently high energy strikes a nucleon
included in an atomic nucleus and produces the boson~Z$^{0} $, then
{\it in statu nascendi\/} this boson moves with respect to the
nucleus (its remainder) with a velocity that sufficiently intensifies
the electromagnetic field. In particular, neutrons present in nuclei
should be struck by neutral particles, while protons by
negatively charged ones. In the case of nuclei so large that we may
speak of peripheral nucleons, the collision with such a nucleon
(``tangent'' collision) is the most effective. Note that the
principal m-tachyon is produced {\it only\/} when the
proton in the~$ ^{2}$H,
and perhaps~$ ^{3}$H, nucleus is appropriately struck. When
designing controlled collisions, we can practically use only
electrons or antiprotons as the striking particles. In all the
mentioned collisions we have $ U \cong 1$ and therefore, by Eq.~(18),
the striking particle and the produced principal tachyon have
practically the same direction of motion, but according to our
theory they may have different senses. In the case of opposite
senses for brevity we shall be speaking about {\it backward
tachyons}, and in the case of the same senses about {\it forward
tachyons}. This nomenclature relates to the principal tachyons
only.

The collisions described above should occur in air showers and can
be realized in or at some high-energy colliders. Let us discuss
these two cases in terms of the {\it laboratory\/} (and thus the
{\it earth\/}) reference frame.

The collisions producing tachyons should occur
in the air showers
initiated by cosmic (primary) particles
of energy of $ \sim\!\!10^{13}$~eV
and greater (events above $ 10^{20}$~eV have been reported~[23]).
Thus our hypothesis justifies air shower experiments
designed to detect tachyons. The time-of-flight measurement
experiments (e.g. described in Refs. [8,24,25]) are obviously
more credible than the experiments described and/or cited in
Refs. [5--7,10,11] and designed only to detect charged particles
preceding the relativistic fronts of air showers, though a
massive-measurement experiment of the latter type performed by
Smith and Standil with the use of detector telescopes [26]
has had great
weight. Tachyon candidates were observed in the time-of-flight
experiments [8,24,25] and in many ``preceded front'' ones
including that described in Ref.~[26], but these unlucky candidates
were sunk in backgrounds and/or statistics. Thus, formally, we have
to consider the results as negative. In the light of our hypothesis,
however,
properly designed experiments with air showers (``poor man's
accelerator''~[25]) are worth repeating, the more so as they are
relatively inexpensive.

Let us note that no forward tachyons can be observed in any air
shower experiment performed in the terrestrial reference frame,
since these tachyons cannot practically precede the shower fronts.
In fact, it is easy to calculate from relations~(16), (19), and from
the relativistic law of
addition of velocities that the forward e-tachyons produced in
collisions with nuclei $ ^{40}$Ar can move in this
reference frame with speeds not greater than
$ \sim\!\!1.0000008c $. In the case of nuclei $ ^{16}$O or $ ^{14}$N, 
or $ ^{2}$H in the case of production of the forward m-tachyons,
the upper speed limit is still lower. On the other hand, some
tachyons accompanying those ``slow'' forward tachyons may travel
considerably faster than light towards the ground.
This is possible provided
that the angle, denoted by $ \psi $ for short, between the motion
directions of such a forward tachyon and of its accompanying tachyon
is sufficiently large.\footnote[11]{In every given reference
frame, if a principal tachyon moves with a speed $ |W| < \infty $
and if the
angle $ \psi $ between the velocity $ W $ and velocity $ V $ of a 
tachyon accompanying this principal one is, for simplicity, smaller
than $ \pi/2 $, then $ |V| \leq c|W|/
[c\cos\!\psi + (W^{2} - c^{2})^{1/2}\sin\!\psi] $
and there is a lower limit for
$ \psi $, namely
$ \arccos (c/|W|) < \psi < \pi/2 $ in the case under
consideration. Of course $ |V| > c $ and $ |W| > c$.} Unfortunately,
these fast accompanying tachyons cannot be observed in typical
``preceded front'' experiments since they escape from the showers
sidewise. They could be observed in the previous
time-of-flight experiments in the cases when the shower axis was
largely inclined with respect to the flight corridor of the
detector (large~$ \psi $).

The described situation seems to explain the poor statistics
obtained from the previous experiments, and suggests how to design
new air shower experiments to search for tachyons. It seems that
the best solution would be an {\it apparatus with many time-of-flight
corridors of various directions}. In order to increase efficiency,
such an apparatus should be possibly close to the region of tachyon
production (mountains? balloons?). To increase credibility, simple
air shower detectors (placed on the ground for convenience) can
additionally be used. They should be far from the main apparatus
(its projection on the ground) to act when $ \psi $ is large, i.e.,
when the registered showers are remote or largely inclined.
If some tachyon flights through the main apparatus coincide
with the signals from some of the additional detectors,
then we get stronger evidence that tachyons are produced in
air showers. The use of the main apparatus alone should also give
us valuable results without detecting any showers.

The appearance of tachyon candidates in some
previous ``preceded front''
experiments can be explained as the arrival of tachyons accompanying
the backward tachyons. The backward tachyons produced in air showers
are slightly faster than $ 5c/3$ in the terrestrial reference frame.
Thus, at sufficiently high altitudes (balloons? satellites?),
they should be easily identified as tachyons.

Failure of the previous air shower experiments may also be explained
by the very low deuterium content (cf. the beginning of this section)
in the earth's atmosphere. Indeed, if the principal e-tachyons do
not exist in nature but the principal m-tachyons do
(cf. the fourth paragraph in Section~3), then the probability of
production of principal tachyons is very low. Then, however, this
probability strongly depends on weather. Roughly speaking, the
cloudier the skies the higher the probability. It seems that
this aspect has not been taken into account in the
experiments performed hitherto. If the principal tachyons are
only the m-tachyons,
then the efficiency of air shower experiments may be
increased by introducing extra deuterium. For instance,
we can place the above mentioned
apparatus (i.e. that with many time-of-flight corridors)
{\it inside\/} a large balloon filled
with hydrogen and next dispatch the balloon to the region of
tachyon production.

In the case of performing tachyon search experiments with the use of
accelerators we can choose the striking particles (practically
either electrons or antiprotons), the nuclei to be struck,
and the energy of collisions. Relations (19) and (20) mean that
the strongest colliders should be employed. At present,
however, we can
only direct a beam of electrons or antiprotons onto a stationary
target. This would give us principal tachyons such as in the case of
air showers, i.e. forward tachyons so ``slow'' that
indistinguishable as tachyons and backward tachyons
slightly faster than $ 5c/3$. As regards accompanying tachyons,
we would have a much better situation since the target can be
surrounded with tachyon detectors, e.g. with time-of-flight ones.
The fact that tachyon candidates were observed in air shower
experiments indicates that there should be no problems with the
range of tachyons in the collider experiments. A collider with
a high energy beam of atomic nuclei would extend our empiric
possibilities. We could then control the observed speeds of
backward and forward tachyons and, in consequence, change the observed
velocities of the accompanying tachyons. Besides, we could then
produce principal m-tachyons (cf. the preceding paragraph), which
is impossible in the near future when a stationary target is used.
For instance, a beam of electrons of energy of $ \sim\!25$~GeV
or a beam of antiprotons of energy of $ \sim\!0.1$~TeV when
colliding with a beam of deuterons of energy of $ \sim\!\!1$~TeV 
($ \sim\!0.5$~TeV/u) or of $ \sim\!0.24$~TeV
($ \sim\!0.12$~TeV/u), respectively,
would already realize the production conditions, whereas in the case
of the deuterium target the energy of the striking negative
particles must be $ \sim\!26$~TeV. When using stationary targets
to produce principal e-tachyons, we need the striking negative
particles of energy of $ \sim\!0.8$~TeV for the targets made of
heavy nuclei, and of $ \sim\!2$~TeV for the targets made of light
nuclei.

Let us note that in the experiments designed to detect tachyons the
existence of a reference frame preferred for the tachyons should be
taken into account.\footnote[12]{The
existence of such a reference frame has been considered or postulated
by many authors. Most of the relevant literature is cited in Refs.
[1,2,27]. Some ideas are, however, in conflict with empiric data,
some others can only be verified by means of tachyons.
According to the latter ideas such a frame is imperceptible for
bradyons and luxons, which means that this frame is a usual
non-preferred inertial reference frame for all the tachyonless
phenomena. This is not contradictory to relativity (which
has been verified only in the bradyonic and luxonic domains) and is
not empirically ruled out since tachyons have not yet been employed.
The most natural idea (i.e. when the (local) Minkowski's spacetime
is assumed to be spatially isotropic also for tachyons) has
thoroughly been analyzed in Section~3 of Ref.~[2]. Following this
idea, many authors suggest that the frame in question is
that in which the cosmic microwave background radiation is
isotropic. If their intuition is correct, then in terrestrial
experiments this frame can be revealed only by means of tachyons
which are very fast (over $ \sim\!800c$) in the laboratory reference
frame. If, however, the ``tachyon corridor'' described by Antippa
and Everett [28,29] did exist, then ``slow'' tachyons would be 
sufficient to reveal it.} In terrestrial experiments we should
therefore analyze the measurements in correlation with the time of
the day, and additionally, in long-lasting experiments, with the 
season of the year. It seems obvious that from this point of
view the experiments with the use of colliders are more suitable
than those with air showers.

\vskip16pt

\centerline{{{\bf\large 6 Comments on tachyonic neutrinos}}}
\centerline{{{\bf\large and on unfortunate terminology}}}

\vglue12pt

\noindent
The results of some experiments from which the neutrino mass is
being squeezed out, astonish physicists for over two decades.
Namely, when the relativistic formulae for conservation of
four-momentum are used, the experimenters obtain ``negative''
values for the squared rest mass of neutrinos. (A good deal of the
literature concerning the muon neutrino
is given in, e.g., Refs. [30,31], and that concerning
the electron neutrino is given in Refs. [32--34].) Two
problems then arise -- physical and terminological.

The squared mass values mentioned above are burdened with
empiric errors so large that the opinion that the neutrinos have
zero mass can still be maintained. A detailed critical
analysis and list of empiric data concerning the electron
neutrino from $ \beta $-decay are given in Ref.~[32]. However, it is
striking that independent experiments systematically give the
``negative squared rest mass'' of neutrinos (which in reality would
be neither negative nor rest mass as we shall see below),
especially in the case of the muon neutrino from $ \pi $-decay,
i.e. from a simple phenomenon. If these results were confirmed,
then, in terms of relativity, such neutrinos would really be
faster than light, and the universe would be filled with almost
noninteracting tachyons.\footnote[13]{A peculiar model of
the universe, according to which some known phenomena are caused
by tachyons, has been proposed by Steyaert [35,36].}

In the tachyonic literature it is frequently stated that ``the
squared rest mass of tachyons is negative'', and consequently some
authors conclude that ``the rest mass of tachyons is imaginary''.
Besides, the sentence ``photons have zero rest mass'' is almost
commonly used. Thus someone may be under the impression that many
authors use relativistic terms and formulae without understanding
their meanings. Let us make a few elementary remarks.

In relativity the term ``rest mass'' does not make sense in the
case of luxons and tachyons, since the state of rest can be
reasonably defined for these objects neither within standard
relativity nor in its consistent extensions. This is obvious in the
luxonic case since, e.g., the Lorentz transformation is singular
for speeds equal to~$ c $. For the tachyonic case see Footnote~6.

As regards the phrases ``negative squared mass'', ``imaginary
mass'', and ``photon's zero mass'', we shall proceed step by
step.

Consider the world line $ x^{\mu}\left(\sigma\right) $ of a
pointlike object. Assume, for simplicity, that the object is free
in flat spacetime endowed with the Lorentzian coordinates (i.e.
$ x^{\mu}\left(\sigma\right) $ is straight), that $ \sigma $ is the
normalized affine parameter of $ x^{\mu}\left(\sigma\right) $, and
that the signature is, e.g., $ + + + - {}$. Note that in the metric
form expressions, $ ds^{2} = dx_{\mu}dx^{\mu},\ ds^{2} $ is only a
conventional symbol, and therefore it need not be the square of an
infinitesimal real quantity. In the case under consideration
$$
ds^{2} = dx^{2} + dy^{2} + dz^{2} - c^{2}dt^{2},\eqno (21)
$$
and for $ x^{\mu}\left(\sigma\right) $ we have
$$
ds^{2} = -k\left(d\sigma\right)^{2},\eqno (22)
$$
where $ d\sigma $ is indeed an infinitesimal real quantity, and
where the discrete dimensionless parameter $ k $ is as follows:

$k = 1$ in the bradyonic (timelike, subluminal) case,

$k = 0$ in the luxonic (null, luminal) case, and

$k = -1$ in the tachyonic (spacelike, superluminal) case.

\noindent (If the signature $ + - - - {}$
were chosen, then by Eq.~(22) we would have $ k = -1 $ in
the bradyonic case and $ k = 1 $ in the tachyonic case.) Dividing
Eqs. (21) and (22) by $
\left(d\sigma\right)^{2} $ we get
$$
-k = \left(u^{x}\right)^{2} + \left(u^{y}\right)^{2} +
\left(u^{z}\right)^{2} - \left(u^{t}\right)^{2},\eqno (23)
$$
where $ u^{\mu} := dx^{\mu}/d\sigma $ is a four-velocity vector.
The kinematical Eq.~(23) concerns every type of world lines --
timelike, null, and spacelike. The type is determined by~$ k $.

Multiplying Eq.~(23) by $ m^{2}c^{2} $, where $ m $ has
the mass dimension (we do {\it not yet\/} determine physical
meanings of~$ m $), we get the well-known special relativistic
formula for a four-momentum vector $ p^{\mu} $:
$$
-km^{2}c^{2} = \left(p^{x}\right)^{2} + \left(p^{y}\right)^{2} +
\left(p^{z}\right)^{2} - \left(p^{t}\right)^{2} \equiv {\bf p}^{2} -
c^{-2}E^{2},\eqno (24)
$$
where
$$
p^{\mu} := mcu^{\mu},\eqno (25)
$$
and where $ \left(p^{x}\right)^{2} + \left(p^{y}\right)^{2}
+ \left(p^{z}\right)^{2} \equiv {\bf p}^{2} $ and $
\left(p^{t}\right)^{2} \equiv c^{-2}E^{2} $. If we had $ m =
0$, then by definition (25) we would have no four-momentum, i.e.
no object on our world line (not speaking of that
the multiplication of
equations by zero does not make sense). Thus
$$
m \neq 0.\eqno (26)
$$
If $ m $ were imaginary, then by definition (25) also the
four-momentum components $ p^{\mu} $ would be imaginary, which
would give us a new physics yet
unknown.\footnote[14]{The first
appearance of imaginary mass in the tachyonic literature is
fairly funny. Namely, some authors have put $v^{2} > c^{2}$ in the
known relativistic formula for energy, $E = mc^{3}\left(c^{2} -
v^{2}\right)^{-1/2}$, which is valid for bradyons and not for
tachyons, and to avoid the imaginary energy (interactions?) they
assumed an imaginary $m$. The tachyonic literature is full of
surprising ideas, including incantations, e.g.,
``pseudo-antiorthogonal transformations'' [37] or the requirement
to use
the term ``pseudo-Riemannian'' with regard to the Riemannian space
with the relativistic signature $ + + + - {}$ or $ + - - - {}$
[37,38]. Some ideas are brilliant, e.g., to use simultaneously two
signatures ($ + + +
- {}$ and $ + - - - {}$) in one description of spacetime
relations~[37]. The largest list of tachyonic publications is given
in Ref.~[37]. Most of them, however, represent the unfortunate trends
(see the beginning of our Section~1, the end of Footnote~6, and
Footnote~15), whereas a number of papers criticizing these trends
is omitted (some of them are cited in Refs. [1,2,19]).}
If we had real $ m < 0$, then by definition
(25) we would have opposite senses of the four-vectors $ u^{\mu} $
and~$ p^{\mu} $. Such a situation is yet unknown and today seems
strange, though perhaps it will be considered in future. Anyway, we
are entitled to put real $ m > 0 $ for {\it every\/} type of the
objects under consideration (Ockham's principle!).

The unfortunate phrases have resulted from the fact that some
authors have not taken into account the existence of three values
of $ k $ (1, 0, $ -1 $) and have applied the bradyonic variants of
Eqs. (21)--(24) for luxons and
tachyons.\footnote[15]{Attempts to
escape trouble in the tachyonic case have consisted in the
confusion between mappings and transformations. This confusion,
frequent in the tachyonic literature, has been discussed in Refs.
[2,18] (in the context of the superluminal reference frame
problem, cf. Footnote~6). Also frequent attempts have consisted in
interchanging the meanings of the energy and momentum terms in the
bradyonic variant of Eq.~(24), without taking into account that
momentum has three components in the physical spacetime.
Effects of such an interchange have been described in Footnote~2 in
Ref.~[3].} The use of proper values of $ k $ allows to avoid the
difficulties. If, for instance, the general formula for energy, $ E
= \left({\bf p}^{2}c^{2} + km^{2}c^{4}\right)^{1/2} $ (for the
signature $ + + + - {}$), had been applied in the mentioned works
on neutrinos, instead of its bradyonic variant $ E = \left({\bf
p}^{2}c^{2} + m^{2}c^{4}\right)^{1/2} $, then the embarrassing
``negative squared rest mass'' would not have appeared; there would
then have been a positive quantity, for $ k
= -1 $ under the assumption that those neutrinos are tachyons. Of
course the term ``rest mass'' would then be improper.

In the bradyonic case, $ m $ is the rest mass of our object. In the
luxonic case the physical meaning of $ m $ is not determined in
general, though it is so for the photon for which $ m = c^{-2}E =
c^{-2}h\nu
> 0$. Anyway, the dynamical luxonic relation $ {\bf p}^{2}c^{2} =
E^{2} $ does {\it not\/}
result from the condition $ m = 0$, which is false (inequality~(26)),
but it does result from the condition $ k = $ 0, i.e. it is
determined at the {\it kinematical\/} level of Eqs. (21)--(23).
In the tachyonic case we have yet no operational definition of $ m
$ (for lack of rest), and therefore the term ``masslike quantity''
has been proposed (cf. Footnote~3).

\vskip16pt

\centerline{{{\bf\large 7 Concluding remarks}}}

\vglue12pt

\noindent
Solution (1)--(7) of the Einstein--Maxwell equations yields a
realistic picture of the tachyonic phenomenon. The existence of this
solution can therefore be regarded as an indication on the part of
general relativity in favour of the tachyon's existence in nature,
considering the analogy to many theoretical predictions that found
later empirical confirmation. The solution is the basis of the
hypothesis presented in this paper.

The hypothesis determines the principal empiric conditions of
tachyon production. These conditions can occur when
nonpositive subatomic particles of high energy strike
atomic nuclei other than the proton. Thus, if our hypothesis is
true, we should expect credible tachyons to appear in
properly designed
experiments with air showers or with the use of the strongest
colliders. In the latter type experiments,
not performed hitherto,
the production of tachyons can be controlled.

\vskip12pt

\noindent{\bf\large Acknowledgement} I wish to thank Bogdan Mielnik
for reading the man\-u\-script and helpful discussions.

\vskip16pt

\begin{bibliography}{}

\centerline{{{\bf\large References}}}

\vglue12pt
 [1] R. Girard, L. Marchildon: {\it Found. Phys.} {\bf 14} (1984) 535

 [2] J.K. Kowalczy\'nski: {\it Int. J. Theor. Phys.} {\bf 23} (1984)
27

 [3] J.K. Kowalczy\'nski: {\it Acta Phys. Pol.} {\bf B20} (1989) 561

 [4] J.K. Kowalczy\'nski: {\it The Tachyon and its Fields}, Polish
Academy of Sciences, Warsaw 1996

 [5] J.R. Prescott: {\it J. Phys.} {\bf G2} (1976) 261

 [6] L.W. Jones: {\it Rev. Mod. Phys.} {\bf 49} (1977) 717

 [7] P.N. Bhat, N.V. Gopalakrishnan, S.K. Gupta, S.C. Tonwar:
{\it J. Phys.} {\bf G5} (1979) L13

 [8] A. Marini, I. Peruzzi, M. Piccolo, F. Ronga, D.M. Chew, R.P. Ely,
T.P. Pun, V. Vuillemin, R. Fries, B. Gobbi, W. Guryn, D.H. Miller,
M.C. Ross, D. Besset, S.J. Freedman, A.M. Litke, J. Napolitano,
T.C. Wang, F.A. Harris, I. Karliner, Sh. Parker, D.E. Yount:
{\it Phys. Rev.} {\bf D26} (1982) 1777

 [9] Particle Data Group: {\it Review of Particle Properties},
{\it Phys. Rev.} {\bf D50} (1994) No. 3--I, p. 1811

[10] R.W. Clay: {\it Aust. J. Phys.} {\bf 41} (1988) 93

[11] R.W. Clay, P.C. Crouch: {\it Nature} {\bf 248} (1974) 28

[12] I. Robinson, A. Trautman: {\it Proc. R. Soc. Lond.} {\bf A265}
(1962) 463

[13] I. Robinson, A. Trautman: {\it Phys. Rev. Lett.} {\bf 4} (1960)
431

[14] D. Kramer, H. Stephani, M. MacCallum, E. Herlt: {\it Exact
Solutions of Einstein's Field Equations},
Cambridge University Press, Cambridge 1980 (or VEB Deutscher Verlag
der Wissenschaften, Berlin 1980; two simultaneous editions), p. 259

[15] J.K. Kowalczy\'nski: {\it J. Math. Phys.} {\bf 26} (1985) 1743

[16] J.K. Kowalczy\'nski: {\it Phys. Lett.} {\bf 74A} (1979) 157

[17] G.C. Debney, R.P. Kerr, A. Schild: {\it J. Math. Phys.} {\bf 10}
(1969) 1842

[18] J.K. Kowalczy\'nski: {\it Acta Phys. Pol.} {\bf B15} (1984) 101

[19] L. Marchildon, A.F. Antippa, A.E. Everett: {\it Phys. Rev.}
{\bf D27} (1983) 1740

[20] G. Boillat: {\it Ann. Inst. Henri Poincar\'e} {\bf A5} (1966)
217

[21] J. Pleba\'nski: {\it Lectures on Non-Linear Electrodynamics},
NORDITA, Co\-pen\-ha\-gen 1970, pp. 67 and 68

[22] A. Sch\"afer: {\it J. Phys.} {\bf G15} (1989) 373

[23] T.K. Gaisser, T. Stanev: {\it Phys. Rev.} {\bf D54} (1996) 122

[24] F. Ashton, H.J. Edwards, G.N. Kelly: {\it Nucl. Instrum.
Methods} {\bf 93} (1971) 349

[25] H. H\"anni, E. Hugentobler: in {\it Tachyons, Monopoles, and
Related Topics}, ed. E. Recami, North-Holland, Amsterdam 1978, p. 61

[26] G.R. Smith, S. Standil: {\it Can. J. Phys.} {\bf 55} (1977) 1280

[27] J. Rembieli\'nski: {\it Int. J. Mod. Phys.} {\bf A12} (1997)
1677

[28] A.F. Antippa, A.E. Everett: {\it Phys. Rev.} {\bf D8} (1973)
2352

[29] A.F. Antippa: {\it Phys. Rev.} {\bf D11} (1975) 724

[30] R. Abela, M. Daum, G.H. Eaton, R. Frosch, B. Jost, P.-R. Kettle,
E. Steiner: {\it Phys. Lett.} {\bf 146B} (1984) 431

[31] K. Assamagan, Ch. Br\"onnimann, M. Daum, H. Forrer, R. Frosch,
P. Gheno, R. Horisberger, M. Janousch, P.-R. Kettle, Th. Spirig,
C. Wigger: {\it Phys. Rev.} {\bf D53} (1996) 6065

[32] E.W. Otten: {\it Nucl. Phys. News} {\bf 5} (1995) No. 1, p. 11

[33] J. Ciborowski: {\it Acta Phys. Pol.} {\bf B29} (1998) 113

[34] J. Ciborowski, J. Rembieli\'nski: {\it Eur. Phys. J.} {\bf C8}
(1999) 157

[35] J.J. Steyaert: in {\it Low Temperature Detectors for Neutrinos
and Dark Matter II}, eds. L. Gonzalez-Mestres and D. Perret-Gallix,
Editions Fronti\`eres, 1988, p. 129

[36] J.J. Steyaert: {\it From Experiment to Theory: an Evidence for
New Physics Far Away from Standard Models, Useful Mainly in
Astrophysics}, preprint, 1999

[37] E. Recami: {\it Riv. Nuovo Cimento} {\bf 9} (1986) No. 6

[38] E. Recami: {\it Int. J. Theor. Phys.} {\bf 25} (1986) 905;
{\bf 26} (1987) 913

\end{bibliography}

\end{document}